\documentclass[twocolumn,notitlepage,showkeys,showpacs,longbibliography,prl]{revtex4-1}
\usepackage{stmaryrd}
\usepackage{amsmath}
\usepackage{amssymb}
\usepackage{graphicx}
\usepackage{dcolumn}
\usepackage{bm}
\usepackage{multirow}
\usepackage[colorlinks,linkcolor=blue,citecolor=blue,urlcolor=blue]{hyperref}
\usepackage{color}
\usepackage{ulem}

\usepackage{array}
\usepackage{booktabs}

\begin{document}

\begin{titlepage}
\title{Unconventional anomalous Hall effect from magnetization parallel to the electric field}

\author{Hengxin Tan}
\author{Yizhou Liu}
\author{Binghai Yan}
\email{binghai.yan@weizmann.ac.il}
\affiliation{Department of Condensed Matter Physics, Weizmann Institute of Science, Rehovot 7610001, Israel}

\date{\today}

\begin{abstract}
In the anomalous Hall effect (AHE), the magnetization, electric field and the Hall current are presumed to be mutually vertical to each other. In this work, we propose an unconventional AHE where the magnetization, the electric field and Hall current stay inside the same plane. Such an AHE is odd under time-reversal and exists even when the magnetization is parallel to the electric field or Hall current, different from the planar Hall effect which is even under time-reversal. Here, we term it parallel anomalous Hall effect (PAHE). We reveal that the PAHE exists when all the point group rotational and reflection symmetries are broken where the Berry curvature field is not necessarily parallel to the magnetization axis. We further demonstrate the PAHE in a ferrimagnetic Weyl semimetal FeCr$_2$Te$_4$. 
\end{abstract}

\maketitle
\draft
\vspace{2mm}
\end{titlepage}
\section{Introduction}
The intrinsic anomalous Hall effect (AHE)\cite{Nagaosa} is established on the Berry phase theory\cite{DiXiao} and provides a powerful probe on the time-reversal breaking and the band topology. In general, it is presumed that three vectors -- magnetization, electric field and the Hall current -- are mutually perpendicular to each other, which is like the conventional Hall effect in an external magnetic field. However, the recent discovery of giant AHE in noncollinear antiferromagnets\cite{Nakatsuji2015,Nayak2016,Chen2014,zhang2017strong} questioned this assumption, in which the net magnetization vanishes. Furthermore, the quantized AHE was theoretically proposed to exist with these three vectors being co-planar in some two-dimensional (2D) films that break all reflection symmetries \cite{PRL111p086802,PRB94p085411,PRB96p241103,PRL121p246401}. It is elusive how this in-plane AHE is generalized to three-dimensional (3D) materials and what symmetry condition is required.

\begin{figure*}[tbp]
\includegraphics[width=1.95\columnwidth]{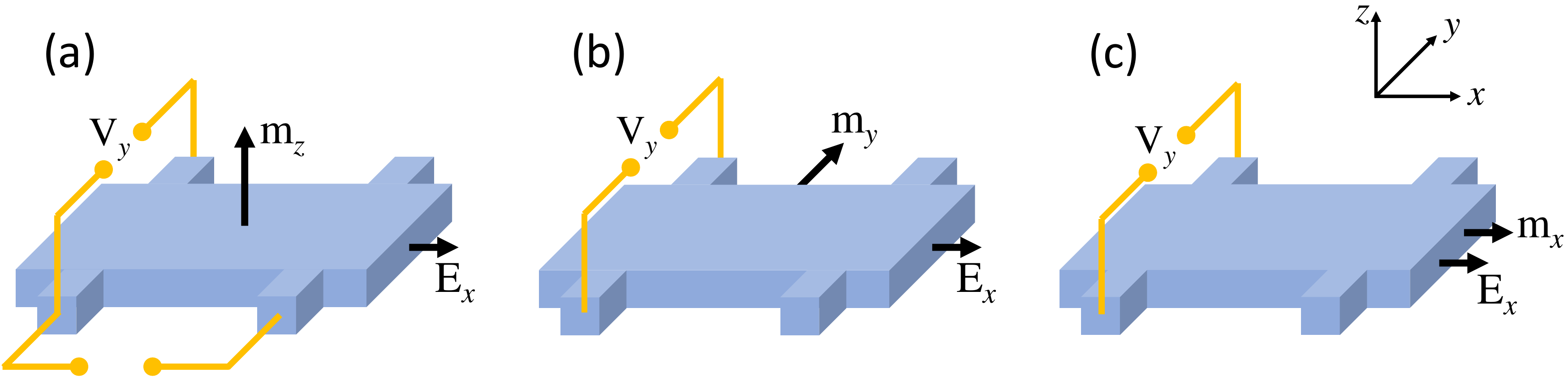}
\caption{\label{Sketche} Schematics of the anomalous Hall effect. (a) The standard Hall bar where the magnetization $\mathbf{m}$ (or magnetic field $\mathbf{H}$) is along $z$, electric field $\mathbf{E}$ is along $x$, and the Hall voltage V$_y$ is measured along $y$. The general electric resistance can be obtained by measuring along $x$. (b) The $\mathbf{m}$ (or $\mathbf{H}$) is along $y$. (c) The $\mathbf{E}$ and $\mathbf{m}$ are in the same direction $x$. The anomalous Hall effect along $y$ ($V_y$) in (b) and (c) are PAHE, which vanishes in the conventional Hall effect and the planar Hall effect. We have to emphasize that the realization of (b) and (c) in real experiment can be achieved by preparing a new sample with the magnetization lying in the sample plane and applying $\mathbf{E}$ parallel to or along with $\mathbf{m}$, instead of rotating $\mathbf{m}$ from (a) directly.
}
\end{figure*}

Here, we express the anomalous Hall current as $\mathbf{J}^{AHE}=\frac{e^2}{\hbar}\mathbf{\Omega}\times \mathbf{E}$, where $\mathbf{\Omega}$ is the total Berry curvature of the band structure, $\mathbf{E}$ the applied electric field and $\frac{e^2}{\hbar}$ the conductance quantum. If any rotational or reflection symmetry exists, both $\mathbf{\Omega}$ and the magnetization $\mathbf{m}$ must lie parallel to the rotational axis or the reflection plane normal, because $\mathbf{\Omega}$ is a pseudo-spin-type vector. Therefore, $\mathbf{J}^{AHE}$, $\mathbf{E}$ and $\mathbf{m}$ are mutually orthogonal in this case. However, if all rotational and reflection symmetries are broken, $\mathbf{\Omega}$ unnecessarily aligns along $\mathbf{m}$. Then the orthogonal relation may get violated. Three vectors,  $\mathbf{J}^{AHE}$, $\mathbf{E}$ and $\mathbf{m}$ can be coplanar and $\mathbf{m}$ may even be parallel to $\mathbf{J}^{AHE}$ or $\mathbf{E}$, as schematically shown in Fig. \ref{Sketche}. We note this unconventional AHE as parallel anomalous Hall effect (PAHE). Like the conventional AHE, PAHE changes sign as reversing the $\mathbf{m}$ direction. It is distinct from the planar Hall effect\cite{Tang2003PRL,Burkov2017PRB,Nandy2017PRL,PhysRevResearch.3.L012006}, which remains the same when flipping the magnetic field and originates in the anisotropic magnetoresistance.

In this work, we investigate the PAHE in general 3D materials. We demonstrate the misalignment between magnetization and Berry curvature if anisotropic spin-orbit coupling (SOC) exits. Unlike the 2D case, we need to break all rotational and reflection symmetries except the spatial inversion to generate PAHE in 3D. We analyze all thirty-two point groups (PGs) and identify the allowed PAHE. Further, we propose the experimental available magnetic Weyl semimetal FeCr$_2$Te$_4$ as a candidate to realize PAHE.

\section{A toy model}
We first illustrate the PAHE by a simple two-band toy model in this section. The two-band anisotropic model Hamiltonian can be written as,
\begin{equation}\label{H}
    \begin{split}
        &H = H_0 + H_Z + H_{SO}, \\
        &H_0  = \frac{\hbar \mathbf{k}^2}{2m^*}, ~~ H_Z = g \mathbf{m} \cdot \boldsymbol{\sigma}, \\
        &H_{SO}  = \lambda_x k_x \sigma_x + \lambda_y k_y \sigma_y + \lambda_z k_z \sigma_z, \\
    \end{split}
\end{equation}
where $H_{SO}$ and $H_Z$ are SOC and Zeeman-like terms, respectively. $\boldsymbol{\lambda} = (\lambda_x, \lambda_y, \lambda_z)$ refers to the SOC strength. $\boldsymbol{\sigma} = (\sigma_x, \sigma_y, \sigma_z)$ is the spin Pauli matrix and  $\mathbf{m} = (m_x, m_y, m_z)$ is the magnetization. If $\lambda_x = \lambda_y = \lambda_z = \lambda$, $H_{SO}$ is reduced to the ordinary isotropic form $\lambda \mathbf{k} \cdot \boldsymbol{\sigma}$.  The Hamiltonian in Eq. \eqref{H} can be rewritten into a compact form $H = d_0 \sigma_0 + \mathbf{d} \cdot \boldsymbol{\sigma}$ ($\sigma_0$ is a 2$\times$2 identity matrix) whose energy band dispersion is given by $\varepsilon_{s\mathbf{k}} = d_0 + s \sqrt{\mathbf{d} \cdot \mathbf{d}}$ ($s=\pm1$) with $d_0 = \hbar^2\mathbf{k}^2/(2m^*)$, and $d_\alpha = \lambda_\alpha k_\alpha + gm_\alpha$ ($\alpha = x,y,z$). The total Berry curvature $\boldsymbol{\Omega}$ can be calculated as
\begin{equation}\label{J_AHE}
\begin{split}
    \boldsymbol{\Omega}_{s\mathbf{k}} =& s \frac{\lambda_x \lambda_y \lambda_z}{2|\mathbf{d}|^3} \left( \mathbf{k} + g\mathbf{m}_\lambda \right) \\
    \boldsymbol{\Omega} =& \sum_{s=\pm} \int \frac{\text{d}\mathbf{k}}{(2\pi)^3} \boldsymbol{\Omega}_{s\mathbf{k}} f_{s\mathbf{k}} \\
    =& \left(-\sum_{s=\pm} \int_{\varepsilon \le \mu} \frac{\text{d}\mathbf{k}}{(2\pi)^3} s\frac{g\lambda_x \lambda _y \lambda_z }{2|\mathbf{d}|^3}\right) \mathbf{m}_\lambda,
\end{split}
\end{equation}
where $\mathbf{m}_\lambda = (\frac{m_x}{\lambda_x}, \frac{m_y}{\lambda_y}, \frac{m_z}{\lambda_z})$. $\mu$ is the chemical potential. Equation \eqref{J_AHE} indicates that the direction of Berry curvature $\boldsymbol{\Omega}$ is parallel to the magnetization $\mathbf{m}$ only in the isotropic case ($\mathbf{m}_\lambda = \mathbf{m}/\lambda$). For generic low-symmetry anisotropic SOC, $\mathbf{m}_\lambda$ or  $\boldsymbol{\Omega}$ is not necessarily collinear with $\mathbf{m}$. 

A similar conclusion also applies to the conventional Hall effect. Based on the Boltzmann transport theory the conventional Hall current ($\mathbf{J}^{HE}$) due to Lorentz force can be derived as 
\begin{equation}
    \mathbf{J}^{HE} = -e^3\tau^2 \int \frac{\text{d}\mathbf{k}}{(2\pi)^3} \mathbf{v} \mathbf{v} \cdot \left[ \mathbf{B} \times \left( \frac{1}{m^*} \right) \mathbf{E} \right] \frac{\partial f_0}{\partial \varepsilon},
\end{equation}
where $\left( \frac{1}{m^*} \right)_{\alpha \beta} = \frac{1}{\hbar^2} \frac{\partial^2 \varepsilon_\mathbf{k}}{\partial k_\alpha \partial k_\beta}$ is the inverse effective mass matrix and $\mathbf{v} = \frac{1}{\hbar} \nabla_\mathbf{k} \varepsilon_\mathbf{k}$ is the group velocity. $f_0$ is the equilibrium Fermi distribution function. In the case of isotropic parabolic band with a constant effective mass \textit{i.e.} $\varepsilon_\mathbf{k} = \frac{\hbar^2\mathbf{k}^2}{2m^*}$, the $\mathbf{J}^{HE}$ can be calculated as $\mathbf{J}^{HE} = \sigma_H \mathbf{B} \times \mathbf{E}$ with $\sigma_H = \frac{e^3\tau^2 n_F}{m^{*2}}$ ($n_F$ is the carrier density at Fermi level). However, in the case of anisotropic effecitive mass model \textit{i.e.} $\varepsilon_\mathbf{k} = \frac{\hbar^2k^2_x}{2m^*_x} + \frac{\hbar^2k^2_y}{2m^*_y} + \frac{\hbar^2k^2_z}{2m^*_z}$, the Hall current becomes $\mathbf{J}^{HE} = \frac{e^3\tau^2 n_F}{m^*_xm^*_ym^*_z} \mathbf{B}_{m^*} \times \mathbf{E}$ with $\mathbf{B}_{m^*} = (m^*_xB_x, m^*_yB_y, m^*_zB_z)$, so that $\mathbf{J}^{HE}$ is not necessarily perpendicular to $\mathbf{B}$.

\begin{table}
\caption{\label{table1} Anomalous Hall conductivity tensor components of three-dimensional point groups that allow PAHE. The tensor is defined as J$_i$ = $\sigma_{ij}$E$_j$ where J$_i$ and E$_j$ are the $i$'th and $j$'th component of Hall current $\mathbf{J}$ and electric field $\mathbf{E}$ respectively. The Cartesian axes $x$, $y$ and $z$ for each crystal system follow the convention used in Ref. \onlinecite{StandardPiezo} and summarized in SM\cite{SM}. The PAHE components of the $D_3$, $C_{3v}$ and $D_{3d}$ PGs depend on the relative positions of the symmetry operations (rotational axis and reflection plane) and Cartesian axes. However, the symmetry restrictions on the tensor are the same. Thus we employ such coordinate sets that one of the in-plane two-fold rotational axes is along $x$ for $D_3$ and $D_{3d}$, and one of the reflection planes is parallel to $xz$ plane for $C_{3v}$. More information can be found in \cite{SM}.}
\renewcommand\arraystretch{1.0}
\begin{ruledtabular}
\begin{tabular}{cccccc}
  Crystal & Point & \multicolumn{3}{c}{Direction of \textbf{m}} \\
  \cline{3-5}
  System   & group & $x$ & $y$ &$z$\\
  \hline
    \multirow{3}{*}{Hexagonal} & $C_{6h}$&$\sigma_{zx}$ &$\sigma_{yz}$ & \\
                               & $C_{3h}$&$\sigma_{zx}$ &$\sigma_{yz}$ & \\
                               & $C_6$   &$\sigma_{zx}$ &$\sigma_{yz}$ & \\
                               \specialrule{0em}{3pt}{3pt}
                               
    \multirow{5}{*}{Trigonal}  & \uline{$D_{3d}$}&  &$\sigma_{xy}$& \\
                               & \uline{$C_{3v}$}&$\sigma_{xy}$ & & \\
                               & \uline{$D_3$}   &  &$\sigma_{xy}$ & \\
                               & $C_{3i}$&$\sigma_{xy}$, $\sigma_{zx}$ &$\sigma_{xy}$, $\sigma_{yz}$ & \\
                               & $C_3$   &$\sigma_{xy}$, $\sigma_{zx}$ &$\sigma_{xy}$, $\sigma_{yz}$ & \\
                               \specialrule{0em}{3pt}{3pt}
    \multirow{3}{*}{Tetragonal}& $C_{4h}$&$\sigma_{zx}$ &$\sigma_{yz}$ & \\
                               & $S_4$   &$\sigma_{zx}$ &$\sigma_{yz}$ & \\
                               & $C_4$   &$\sigma_{zx}$ &$\sigma_{yz}$ & \\
                               \specialrule{0em}{3pt}{3pt}
    \multirow{3}{*}{Monoclinic}& $C_{2h}$&$\sigma_{xy}$ &  &$\sigma_{yz}$\\
                               & $C_s$   &$\sigma_{xy}$ &  &$\sigma_{yz}$\\
                               & $C_2$   &$\sigma_{xy}$ &  &$\sigma_{yz}$\\
                               \specialrule{0em}{3pt}{3pt}
    \multirow{2}{*}{Triclinic} & $C_i$   &$\sigma_{xy}$, $\sigma_{zx}$ & $\sigma_{xy}$, $\sigma_{yz}$ & $\sigma_{yz}$, $\sigma_{zx}$ \\
                               & $C_1$   &$\sigma_{xy}$, $\sigma_{zx}$ & $\sigma_{xy}$, $\sigma_{yz}$ & $\sigma_{yz}$, $\sigma_{zx}$ \\
\end{tabular}
\end{ruledtabular}
\end{table}

\section{Symmetry restrictions on anomalous Hall effect}
Our above discussions indicate that symmetry is a decisive factor for the appearance of PAHE. Let's now consider from a generic aspect how the symmetry restricts the AHE. In the experiment, the AHE is generally measured with magnetization $\mathbf{m}$ or magnetic field $\mathbf{H}$ along the high-symmetry directions. For this reason, and also for the sake of simplicity, we start from the thirty-two PGs and apply the magnetization $\mathbf{m}$ along the three orthogonal axes of the Cartesian coordinate systems conventionally used for all crystal systems \cite{StandardPiezo}, as summarized in SM\cite{SM}. Note that for the trigonal crystal system, we employ the hexagonal lattice settings. The elements of the anomalous Hall conductivity (AHC) tensor $\sigma$ that allow PAHE are summarized in Table \ref{table1} after considering the symmetry broken by $\mathbf{m}$ (more details can be found in the SM\cite{SM}). The PAHE is not allowed for (i) the five cubic PGs, (ii) the dihedral PGs except for $D_{3}$ and $D_{3d}$ and (iii) the $C_{nv}$ PGs except for $C_{3v}$. One common feature of these PGs is that each of the three Cartesian axes shows at least one of the $n$-fold rotational ($n \ge 2$) and reflection symmetries upon applying the $\mathbf{m}$. For example for $C_{4v}$, the reflection $\mathcal{M}_x$ (reflection plane $yz$) and $\mathcal{M}_y$ (reflection plane $xz$) are maintained respectively with $\mathbf{m}$ along $x$ (\textit{i.e.} [100] direction) and $y$ (\textit{i.e.} [010] direction). For $D_{3h}$ the two-fold rotational symmetry along the in-plane lattice vector $\bm{a}$ (\textit{i.e.} $x$) is preserved with $\mathbf{m}$ along $x$ and the reflection symmetry $\mathcal{M}_y$ (reflection plane $xz$) is preserved with $\mathbf{m}$ along $y$. The synergy of these remaining rotational/reflection symmetries prohibits the PAHE in the PGs mentioned above. However, if there is at least one axis that shows no rotational and reflection symmetry upon applying $\mathbf{m}$, the PAHE is allowed, which comprise Table \ref{table1}. Let's take $D_3$ as an example. We take one of the two-fold rotational axes along the $x$ direction (\textit{i.e.}, along in-plane lattice vector $\bm{a}$ \cite{SM}). No rotational symmetry is maintained when $\mathbf{m}$ is along $y$, but the two-fold rotational symmetry along $x$ is maintained when $\mathbf{m}$ is along $x$. As a result, PAHE is allowed with $\mathbf{m}$ along $y$ in $D_3$.

The above discussions establish for PAHE the preconditions of breaking both rotational and reflection symmetries. But it is still not clear about the plane in which the PAHE appears. By analyzing the symmetries upon applying $\mathbf{m}$, we find that, while the reflection (rotational) symmetry is broken when $\mathbf{m}$ is in the reflection plane (the plane perpendicular to the rotational axis), the combination of reflection $\mathcal{M}$ (two-fold rotation $\mathcal{C}_2$) and time-reversal symmetry $\mathcal{T}$, \textit{i.e.} $\mathcal{M} \mathcal{T}$ ($\mathcal{C}_2 \mathcal{T}$), is maintained. Such combinations of $\mathcal{M} \mathcal{T}$ and $\mathcal{C}_2 \mathcal{T}$ import additional restriction for the AHE. Let's consider the most interesting in-plane component $\sigma_{xy}$ (similar discussion below can be applied to the other components). For example for $C_{3h}$ PG as shown in Table \ref{table1}, the PAHE is only allowed when $\mathbf{m}$ lies in the $xy$ plane. However, due to the maintained $\mathcal{M} \mathcal{T}$ ($\mathcal{M}$ here is the reflection $\sigma_h$ along $z$ axis), the AHE in $xy$ plane ($\sigma_{xy}$) is forbidden. For $C_n$ PG ($n$ is even), now the $\mathcal{C}_2\mathcal{T}$ plays the role of $\mathcal{M} \mathcal{T}$ in $C_{3h}$. Similar discussions can be performed for other PGs. In short, the in-plane AHE is not allowed in the PGs with either even-fold rotational symmetry (the rotational axis is along $z$ direction) or $\sigma_h$ reflection when $\mathbf{m}$ lies in the $xy$ plane.

Now we discuss the two-dimensional case. Previous works\cite{PRL111p086802,PRB94p085411,PRB96p241103,PRL121p246401} revealed the possibility to realize in-plane quantized AHE effect with in-plane magnetization on the precondition of breaking all the reflection symmetries. However, rotational symmetry is no well considered before. This may be because that the materials or models considered therein have no two-fold rotational and reflection symmetries along the out-of-plane direction, thus do not suffer from $\mathcal{C}_2\mathcal{T}$ or $\mathcal{M} \mathcal{T}$ with in-plane magnetization. Here we emphasize the role played by $\mathcal{C}_2\mathcal{T}$ and $\mathcal{M} \mathcal{T}$ and conclude that the PAHE can never happen in $pure$ 2D case with in-plane magnetization (regardless of the stability of such a magnetization). But the PAHE  can happen with in-plane magnetization in the 2D PGs of $C_1$/$C_3$ and $D_1$/$D_3$ with the prerequisite of breaking $\sigma_h$ reflection, \textit{e.g.}, by a slight buckling of the plane.

While the above discussions for the 2D case with in-plane magnetization are robust, we have to emphasize the arguments for 3D PGs are based on the assumption that the magnetization $\mathbf{m}$ is along the specified high-symmetry directions. If $\mathbf{m}$ is misaligned to the high-symmetry direction that destroys all the rotational and reflection symmetries as well as the combinations $\mathcal{M} \mathcal{T}$ and $\mathcal{C}_2 \mathcal{T}$, then PAHE can show up in any crystal class. Such symmetry breaking process can be realized by either an external magnetic field $\mathbf{H}$ along a general direction or a particular type of intrinsic magnetization. Besides, though based on the magnetization from the ferromagnetic configuration, our symmetry considerations are also applicable to other magnetic configurations such as anti-ferromagnetic configuration and non-colinear magnetic configuration.

\begin{figure}[tbp]
\includegraphics[width=0.99\columnwidth]{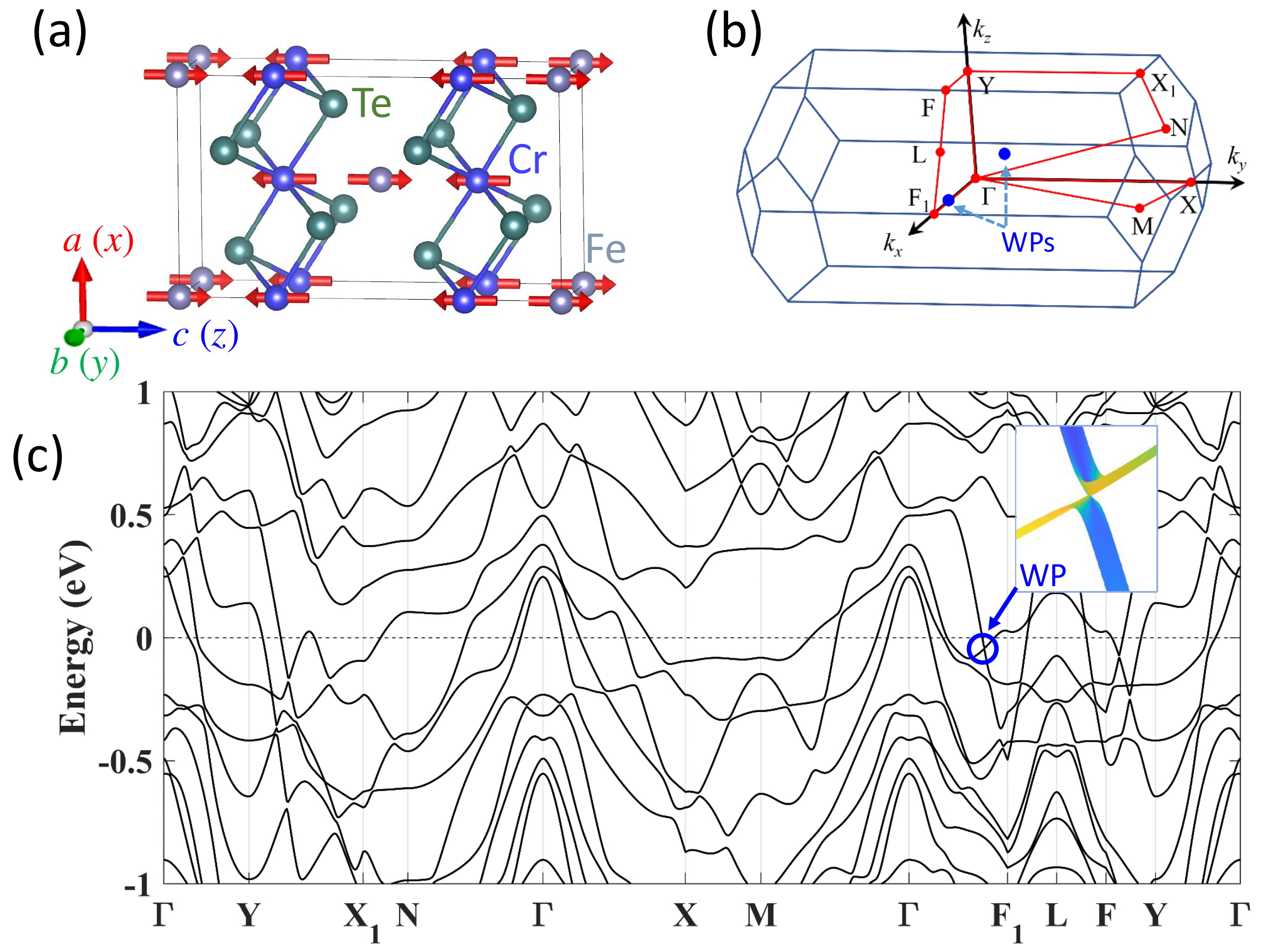}
\caption{\label{band} Crystal and band structures of FeCr$_2$Te$_4$. (a) Crystal and magnetic structure of FeCr$_2$Te$_4$ with the red vectors on Cr and Fe atoms showing the experimentally confirmed ferrimagnetic configuration, where the local moments are aligned ferromagnetically along $\bm{c}$-axis in each sublattice but antiferromagnetically aligned between the two sublattices. (b) The first Brillouin zone of the primitive cell together with the high-symmetry points used in (c) are shown. The blue dots represent a pair of Weyl points near the Fermi level. (c) The band structure under the experimental ferrimagnetic configuration including SOC (the energy is referenced to the Fermi level). The band crossing on $\Gamma$F$_1$ at the energy of $-$44 meV forms a type-I Weyl point (WP) and the inset shows the dispersion relation near the  WP on the $k_x$-$k_y$ plane. Yellow (Blue) color stands for higher (lower) band weight of Fe. }
\end{figure}

\section{Material prediction}
We now demonstrate by first-principles calculation the PAHE in the ferrimagnetic Weyl semimetal FeCr$_2$Te$_4$, which has been reported recently about the ferrimagnetism and general AHE from experiment\cite{PRB102p085158,PRB103p045106}. The crystal structure of FeCr$_2$Te$_4$, as shown in Fig. \ref{band}(a), can be regarded as the Fe-intercalated AA-stacking of 1T phase of a transition-metal dichalcogenide (CrTe$_2$) with distortions. It has a monoclinic structure with a space group of $I$2/m ($C_{2h}$), and the angle between the lattice vectors $\bm{a}$ and $\bm{c}$ is 90.01$^\circ$ (the small deviation from 90$^\circ$ will be ignored below). The two-fold rotational symmetry $\mathcal{C}_2$ is along lattice vector $\bm{b}$ and reflection $\mathcal{M}$ is with respect to $\bm{ac}$ plane. The experimentally confirmed ferrimagnetic configuration with the easy axis along $\bm{c}$ is shown in Fig. \ref{band}(a). Within such a magnetic configuration, both the $\mathcal{C}_2$ and $\mathcal{M}$ are broken, but the combined $\mathcal{C}_2 \mathcal{T}$ and $\mathcal{M} \mathcal{T}$ symmetries are maintained.

The Perdew-Burke-Ernzerhof \cite{PRL77p3865} level band structure, as calculated by the Density Functional Theory as implemented in Vienna $ab$ initio simulation package \cite{vaspPRB,KRESSE199615}, is shown in Fig. \ref{band}(c) where SOC is included. FeCr$_2$Te$_4$ is metallic which is different from its isostructural FeCr$_2$Se$_4$, who is an antiferromagnetic insulator \cite{PRB62p10185, NJP10p055014}. There are many band crossing points in the band structure, some of which are potential Weyl points (WPs). For example, by employing WannierTools software \cite{WannierTools}, we identify that the band crossing point on $\Gamma$F$_1$ at $-$44 meV is a type-I WP as shown in Fig. \ref{band}(c). Due to the maintained inversion symmetry, there is another WP corresponding to the one on $\Gamma$F$_1$. The positions of this pair of WPs in the Brillouin zone are shown in Fig. \ref{band}(b) by blue dots. Such WPs (and other band crossing points) generally contribute much to the Berry curvature and thus to the intrinsic AHC who is the integral of Berry curvature over all the occupied bands. But be aware that the WP is not a prerequisite of AHE according to the symmetry analyses above.

\begin{figure}[tbp]
\includegraphics[width=0.9\columnwidth]{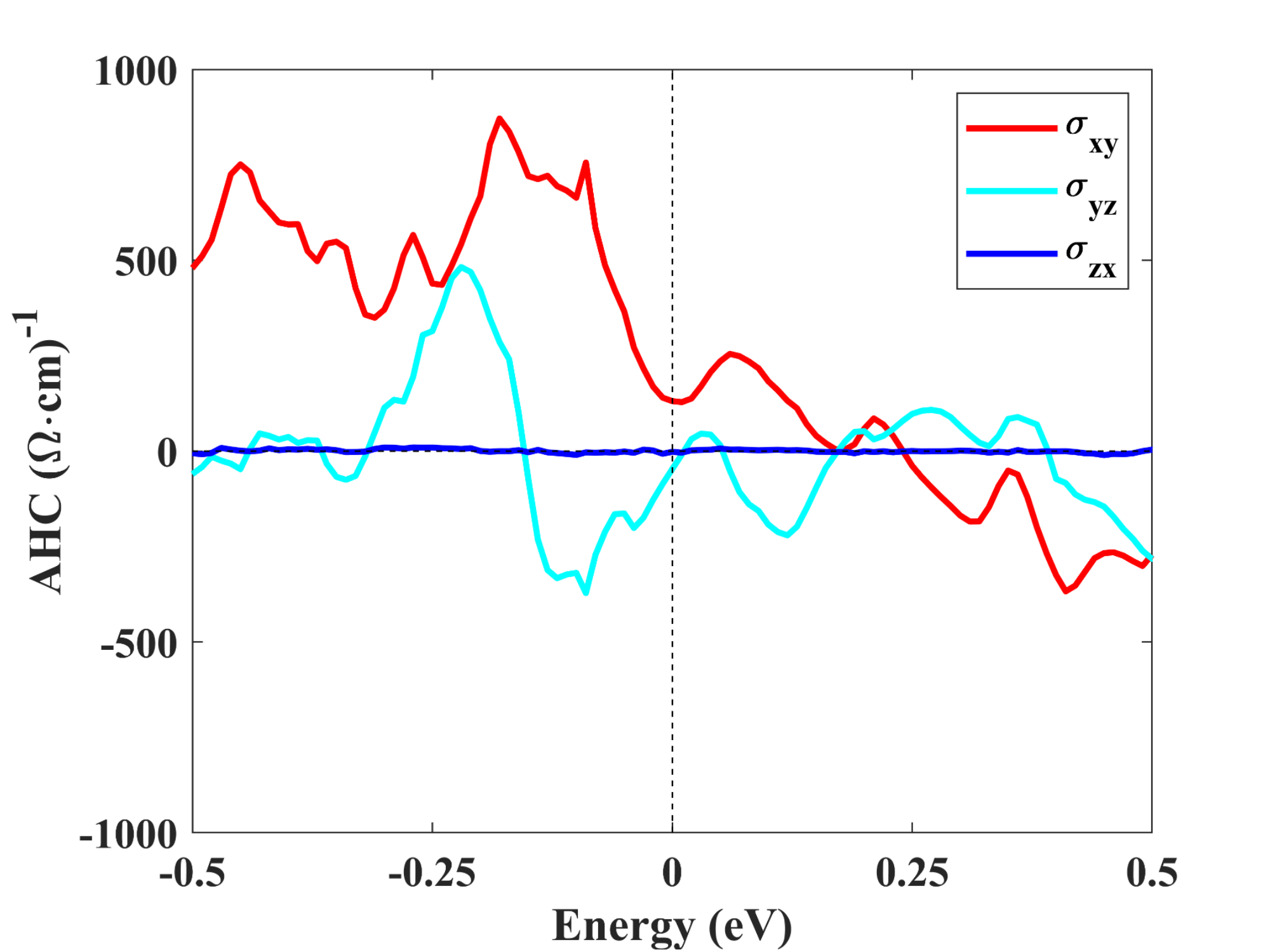}
\caption{\label{ahc} Anomalous Hall conductivity (AHC) of FeCr$_2$Te$_4$ under the experimental ferrimagnetic configuration.
The energy is relative to the Fermi energy.}
\end{figure}

Based on a tight-binding Hamiltonian as obtained with the maximally localized Wannier functions\cite{MLWF}, we calculate the intrinsic AHC by using the Kubo formula approach. Figure \ref{ahc} shows the calculated AHC under the experimental ferrimagnetic configuration. The component $\sigma_{xy}$ (magnetization along $z$) which represents the general AHE, shows a value of about 130 ($\Omega\cdot$cm)$^{-1}$ at the Fermi energy. This intrinsic AHC is larger than the experimental value where the extrinsic contribution dominates the AHE of FeCr$_2$Te$_4$ as discussed in literature \cite{PRB103p045106}. Here we focus on the intrinsic part. While the experiment had reported the $\sigma_{xy}$, the PAHE has not been reported. According to our calculation, the most intriguing component $\sigma_{yz}$, where the electric field or the Hall current is in the same direction of the intrinsic magnetization (\textit{i.e.} $z$ direction), is non-zero, which confirms our above proposal of the PAHE as depicted in Fig. \ref{Sketche}(b) and (c). Notice that the WP shown in Fig. \ref{band}(c) is purely an unexpected surprise and it is not a requirement of PAHE ($\sigma_{yz}$). The value of $\sigma_{yz}$ at the Fermi energy is $\sim$ 50 ($\Omega\cdot$cm)$^{-1}$, in the same order of $\sigma_{xy}$. If the system is slightly doped by hole, $\sigma_{yz}$ can even reach a value of as large as 500 ($\Omega\cdot$cm)$^{-1}$. The $\sigma_{zx}$ is always zero because of the maintained $\mathcal{C}_2 \mathcal{T}$ and $\mathcal{M} \mathcal{T}$ symmetries.

\section{Experiment signature}
Very recently, the in-plane AHE was reported in the potential Dirac or Weyl semimetal material ZrTe$_5$ \cite{ge2020unconventional} when the in-plane magnetic field $\mathbf{H}$ is parallel and perpendicular to the electric field $\mathbf{E}$. ZrTe$_5$ has a PG of $D_{2h}$. According to the symmetry analyses above, this PG does not show PAHE when the $\mathbf{H}$ is along any of the three crystallographic directions since there are always a reflection symmetry and a two-fold rotational symmetry left. However, in the experiment the electrodes are misaligned with the in-plane lattice vectors $\bm{a}$ as manifested in the literature. This misalignment leads to the misalignment between $\mathbf{H}$ and the in-plane $\bm{a}$ or $\bm{c}$ axes when $\mathbf{H}$ is parallel or perpendicular to $\mathbf{E}$, and thus breaks all the symmetry restrictions for PAHE we proposed above. Thus we think the antisymmetric part of the measured unconventional AHE is a signal of PAHE. This actually goes to the proposal above where $\mathbf{H}$ is applied along a general direction for realizing PAHE.

\section{Conclusion}
We have explored the possibility of realizing an unconventional anomalous Hall effect---parallel anomalous Hall effect---where the magnetization (or magnetic moment) is coplanar with the electric field and the Hall current. By symmetry analyses, we reveal that breaking the rotational and reflection symmetries is critical for realizing parallel anomalous Hall effect in three-dimensional. For two-dimensional cases, in addition to the above prerequisites, breaking the additional combinations of two-fold rotational symmetry $\mathcal{C}_2$, reflection symmetry $\mathcal{M}$ (both along out-of-plane direction) and time-reversal symmetry $\mathcal{T}$ (\textit{i.e.} $\mathcal{C}_2 \mathcal{T}$ and $\mathcal{M} \mathcal{T}$) is also essential. By first-principles calculation, we demonstrate this unconventional anomalous Hall effect in a realistic ferrimagnetic Weyl semimetal. Our symmetry discussions also apply to the conventional Hall effect.

\section{Acknowledgements}
We thank Cedomir Petrovic and Daniel Kaplan for inspiring discussions. B.Y. acknowledges the financial support by the Willner Family Leadership Institute for the Weizmann Institute of Science, the Benoziyo Endowment Fund for the Advancement of Science, Ruth and Herman Albert Scholars Program for New Scientists, the European Research Council (ERC) under the European Union's Horizon 2020 research and innovation programme (Grant No. 815869).

%

\end{document}